\documentclass[twocolumn,showpacs,aps,prb]{revtex4}
\usepackage{amssymb}

\usepackage{bm}
\usepackage{times}
\usepackage{amsmath}
\usepackage[dvips]{graphicx}

\newcommand{\smeq}{\! = \!}

\newcommand{\smpl}{\! + \!}
\newcommand{\smmi}{\! - \!}

\newcommand{\be}{\begin{equation}}
\newcommand{\ee}{\end{equation}}
\newcommand{\bea}{\begin{eqnarray}}
\newcommand{\eea}{\end{eqnarray}}

\newcommand{\ci}{\mathrm{i}}

\begin{document}
\title{Out of equilibrium transport through an Anderson impurity: Probing scaling laws within the equation of motion approach}
\author{C. A. Balseiro}
\affiliation{Centro At{\'o}mico Bariloche and Instituto Balseiro, Comisi\'on Nacional de Energ\'{\i}a At\'omica, 8400 S. C. de Bariloche, Argentina}
\affiliation{Consejo Nacional de Investigaciones Cient\'{\i}ficas y T\'ecnicas (CONICET), Argentina}
\author{G. Usaj}
\affiliation{Centro At{\'o}mico Bariloche and Instituto Balseiro, Comisi\'on Nacional de Energ\'{\i}a At\'omica, 8400 S. C. de Bariloche, Argentina}
 \affiliation{Consejo Nacional de Investigaciones Cient\'{\i}ficas y T\'ecnicas (CONICET), Argentina}
\author{M. J. S\'anchez}
\affiliation{Centro At{\'o}mico Bariloche and Instituto Balseiro, Comisi\'on Nacional de Energ\'{\i}a At\'omica, 8400 S. C. de Bariloche, Argentina}
 \affiliation{Consejo Nacional de Investigaciones Cient\'{\i}ficas y T\'ecnicas (CONICET), Argentina}

\begin{abstract}
We study non-equilibrium electron transport through a quantum impurity coupled to metallic 
leads  using the equation of motion technique at finite temperature $T$. Assuming that the
interactions are taking place solely in the impurity and focusing in the 
infinite Hubbard limit, we compute the out of equilibrium density of states
and the differential conductance $G_2(T,V)$ to test several scaling laws.
We find that   ${G_2(T,V)}/{G_2(T,0)}$ is a universal function of
$\textit{both}$ $eV/T_K$ \textit{and} $T/T_K$, being $T_K$ the Kondo 
temperature. The effect of  an in plane magnetic field
on the splitting of the zero bias anomaly in the differential 
conductance is also analyzed. For a  Zeeman splitting  $\Delta$, the  computed differential conductance 
peak splitting  depends only on $\Delta/T_K$, and for large 
fields  approaches the value of $2\Delta $. Besides the traditional two leads setup, we also consider
 other configurations that mimics  recent experiments, namely, an 
impurity embedded in a mesoscopic wire and the presence of a third weakly coupled lead. In these cases, a double peak structure of the Kondo resonance is clearly obtained in the differential conductance while the amplitude of the highest peak is shown to decrease as $\ln(eV/T_K)$. Several features of these results are in qualitative agreement with recent experimental 
observations  reported  on quantum dots.
\end{abstract}
\date{March 17, 2010}
\maketitle

\section{\label{intro}Introduction}

The Kondo effect has been experimentally tested in several devices, ranging
from atoms and molecules in contact with metallic reservoirs to quantum dots
embedded in complex mesoscopic structures. \cite{Nygard2000,DeFranceschi2002,Liang2002,Leturcq2005,Grobis2008,Scott2009}
This many body effect occurs when the ground state of the atom, molecule or
quantum dot (from hereon `the impurity') is degenerated--- typically due to
spin degeneracy. While at high temperature the impurity behaves as a free
spin, correlations build up as the temperature $T$ is lowered. As a consequence,
below a characteristic temperature $T_{K}$, called the Kondo
temperature, the impurity's spin is screened by the spins of the electrons in the
host structure. As a result, the Fermi liquid picture is recovered as $%
T\!\rightarrow \!0$. For a
system in equilibrium,  a characteristic feature of the Kondo regime is the emergence of a 
sharp resonance in the impurity
spectral density at the Fermi level.\cite{Kondo1964,Costi1994,Pustilnik2004} 
In  the absence of external magnetic
fields, $T_{K}$  is the only low energy scale and all  the (zero bias) transport  properties obey a  universal scaling as a
function of $T/T_{K}$.
This effect, which is theoretically well understood, has been reported in a variety of experiments.
\cite{Goldhaber-Gordon1998,Cronenwett1998,vanderWiel2000,Yu2004}

External magnetic fields $B$ or bias voltages $V$
introduce new energy scales that could be of the order of $T_{K}$. In these
conditions, several recent works focused on the search of scaling laws
to test whether the non-equilibrium Kondo physics remains universal.
In particular, measurements of the two terminal differential
conductance $G_2(T,V)$  in single channel quantum dots seem to be
well described, at relatively low  $V$ and low $T$, by a universal scaling
law characterized by two parameters.\cite{Grobis2008}
The universal character of the magnetic field induced splitting of the zero bias anomaly in the differential conductance, has 
also been  tested in a series of experiments.\cite{Kogan2004,Amasha2005,Quay2007} At low $B$, the splitting  decreases 
with $T_{K}$ and the data are compatible
with the predicted universal behavior, being  $B/T_{K}$ the scaling variable.\cite{Moore2000,Logan2001}
A recent publication,\cite{Liu2009} however, reported that at high $B$ there is a crossover regime in which the 
 behavior of the splitting on $T_{K}$  seems to be inconsistent  with the universal character.

From the theoretical point of view, the main ingredients required to compute
the electronic current and related quantities are the retarded and lesser
Green functions \cite{Keldysh1965} of an interacting quantum impurity embedded in a metallic  
network.\cite{Hershfield1992,Meir1992} Much effort has been devoted in recent years to
find reliable approximations for these functions employing different
techniques. 
Most of the  approaches consider the single impurity Anderson
model, in which a single correlated electronic state  is connected to two leads. 
For this configuration, some theoretical
considerations predict an universal scaling of $G_2(T,V)$ in temperature and
applied bias but disagree about the coefficients and parameters that
determine the scaling function.\cite{Costi1994,Ralph1994,Oguri2001} 
Other studies that employ non-equilibrium renormalized perturbation
theory,\cite{Aligia2006} need to include valence fluctuations in order to reproduce the 
simple scaling formula for $G_2(T,V)$. \cite{Rincon2009}
The non-crossing approximation
(NCA) \cite{Meir1993,Wingreen1994} predicts the splitting of the out of equilibrium Kondo
resonance into two peaks, but fails to reproduce other aspects of the Kondo
correlations and the spectral density presents an unphysical peak at the
Fermi level in the presence of a magnetic field. Discrepancies in the
results may be due to the fact that these approaches, although well tested
at equilibrium, have several weak points when extended to the out of
equilibrium situation.\cite{Paaske2004}

Using the equation of motion (EOM) technique with the decoupling scheme proposed 
by Lacroix,\cite{lacroix1981}  the authors
of Refs. [\onlinecite{Entin-Wohlman2005,Kashcheyevs2006}] have recently  obtained  an
explicit analytical expression for the impurity's retarded Green function  in  a general case
in which the impurity is embedded in a  mesoscopic structure at equilibrium.
These studies showed that it is crucial to include in the EOM all the
correlations emerging from the truncated scheme in order to reproduce the
low temperature Kondo correlations that other approximations like the NCA
fail to reproduce.

\begin{figure}[t]
\centerline{\includegraphics[width=8cm]{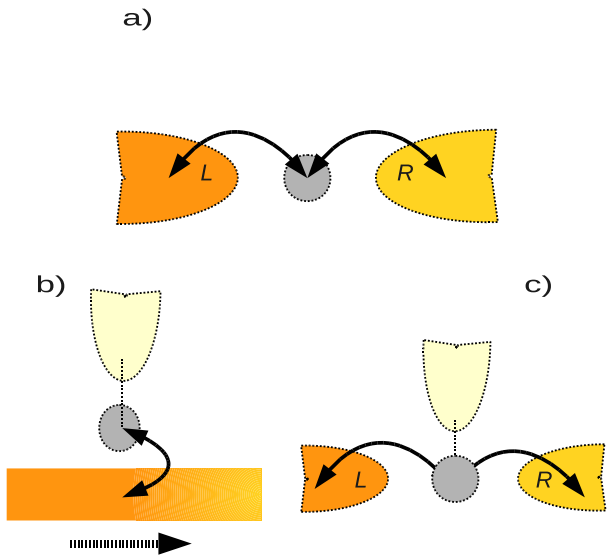}}
\caption{(Color online) a) Schematics of the two terminal quantum dot setup;
b) quantum dot coupled to a biased ballistic wire and a non-invasive probe;
c) quantum dot in a three terminal setup.}
\label{fig1}
\end{figure}

The purpose of the present work is to extend this approach to the out of equilibrium situation to analyze the
transport properties of a single quantum impurity and test some relevant
scaling laws.
First, we shall revisit the conventional two leads setup displayed in Fig.%
\ref{fig1}a) and show that the results for $G_2(T,V)$ are indeed universal; they depend only on $eV/T_K$ and $T/T_K$. We also scale our results with the simple empirical (and approximated) law recently used in the literature \cite{Grobis2008} to scale the experimental data corresponding to a single channel quantum dot  and find a similar behavior. In addition, we discuss the effect of an in-plane magnetic field $B$ on the splitting of the zero bias anomaly in the differential conductance. In agreement with Ref. [\onlinecite{Quay2007}], and in contrast to Ref. [\onlinecite{Liu2009}], we find that the magnitude of the splitting, $2\delta V$, is limited by the Zeeman energy, $\delta V\leq g\mu_B B$.

For single level impurities $G_2(T,V)$ is essentially determined by the impurity's spectral density
$\rho_d(\omega)$. However, the conductance  does
not provide enough information to extract the $V$ dependence of $\rho_d(\omega)$. 
For this purpose, alternative set ups that include an additional lead 
 have been theoretically proposed. \cite{Sun2001,Lebanon2001,Shah2006,Tolea2009} 
Conductance through the additional weakly coupled lead provides a direct probe of the out of equilibrium  spectral
density in the Kondo regime. Closely related experimental setups have been
designed in order to achieve that. In Ref. [\onlinecite{DeFranceschi2002}] a quantum dot
was coupled to a mesoscopic wire formed in a 2DEG that was brought out of
equilibrium by an applied bias voltage. More recently, in Ref. [\onlinecite{Leturcq2005}] the Kondo effect was measured in a ring-shaped dot that
was connected to two strongly and one weakly coupled leads.
Both experiments demonstrated unambiguously the
splitting of the zero bias conductance peak, attributed to the splitting of
the Kondo spectral density due to the double step distribution function of
the carriers in the leads.\cite{Pothier1997}

We shall discuss two related setups (Figs. \ref{fig1}b) and \ref
{fig1}c)) that allow the detection of the out of equilibrium spectral
density of the quantum impurity: i) in Fig.\ref{fig1}b) we mimics 
the set up  used in Ref.[\onlinecite{DeFranceschi2002}] where a quantum dot is
coupled to an out of equilibrium (biased) ballistic wire and a third contact
is used as a weakly coupled (non-invasive) probe; ii) in Fig.\ref{fig1} b)
an extra lead has been added to the two lead configuration reproducing the
setup of Ref.[\onlinecite{Leturcq2005}].

The paper is organized as follows:
Section \ref{model} starts with a description of the model and the EOM approximation.
 In sections \ref{2t} and \ref{3t}  we present the results for the two and three terminal configurations, respectively. 
Section \ref{summary} contains a summary and conclusions. An appendix with some details of the out of equilibrium calculation is also included.

\section{The Model and General Formulation\label{model}}

The Hamiltonian for a single level impurity coupled to a mesoscopic
structure as the ones depicted in Fig.\ref{fig1}, is given by $H=H_{d}+H_{r}$. The  first term in $H$ describes the
isolated impurity, 
\begin{equation}
H_{d}=\sum_{\sigma}\varepsilon_{d\sigma }n_{d\sigma
}+U\,n_{d\uparrow}n_{d\downarrow}\;,
\label{H}
\end{equation}
$n_{d\sigma }=d_{\sigma }^{\dagger }d_{\sigma }^{}$ is the occupation
number operator and  $d_{\sigma }^{\dagger }$ creates an electron with spin
projection $\sigma $ at the impurity level with energy $\varepsilon_{d\sigma
}=\varepsilon_{d}-\sigma g\mu_{B}B/2$ where $\varepsilon_{d}$ is the single particle energy in the absence of
the external magnetic field  $B$.
The last term in Eq. (\ref{H}) describes the Coulomb interaction $U$. In the following we
take the limit $U\!\rightarrow\!\infty $, which forbids the doubly
occupancy of the dot level.
\begin{figure}[t]
\includegraphics[width=.45\textwidth,clip]{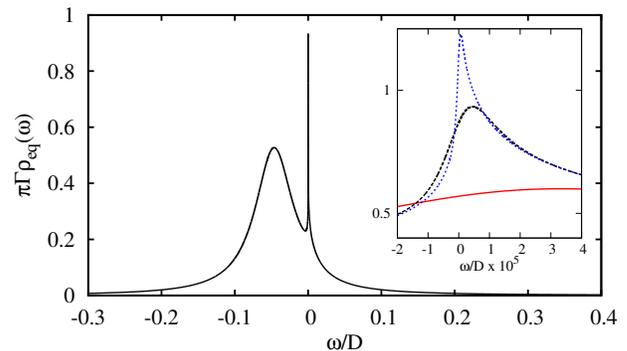}
\caption{(Color online) Equilibrium spectral density for $ \varepsilon%
_d=-4\Gamma$, $\Gamma=0.015D$ y $T=T_K$. Inset: Detailed behavior of the
Kondo peak for different temperature values, $T=0.1,1,10 T_K$}
\label{Ad-E}
\end{figure}
The other term $H_{r}$ describes the non-interacting leads and its coupling
to the impurity, 
\begin{equation}
H_{r}=\sum_{k,\sigma ,\alpha }\varepsilon _{k\sigma }c_{\alpha k\sigma
}^{\dagger }c_{\alpha k\sigma }+(V_{\alpha k}c_{\alpha k\sigma }^{\dagger
}d_{\sigma }+\mathrm{h.c.}),
\end{equation}
where $c_{\alpha k\sigma }^{\dagger }$ creates an electron with momentum $k$
and spin $\sigma $ in the $\alpha$-lead. The leads are in contact to reservoirs at different chemical
potentials $\mu _{\alpha }$ that bring the system out of equilibrium.
In the case of the two lead
configuration depicted in Fig.\ref{fig1}(a), the label $\alpha$ stands for
the left ($L$) and right ($R$) contacts with chemical potentials $\mu _{L}$
and $\mu _{R}$, respectively. In  this configuration, the current $I$
flowing through the impurity can be expressed in terms of the impurity's spectral density
 \cite{Meir1992} 
\begin{equation}
\rho_{d}(\omega )\equiv -\frac{1}{\pi}\sum_\sigma \mathrm{Im} G_{d\sigma
}^{r}(\omega )
\end{equation}
where $G_{d\sigma }^{r}(\omega )$ is the retarded Green function of the impurity.
Using the EOM technique (see the Appendix for the details) we obtain the
following expression for the non-equilibrium (retarded/advanced) Green function, 
\begin{equation}
G_{d\sigma }^{r/a}(\omega )\!=\!\frac{1-\langle n_{\bar{\sigma }}\rangle
\!-\!P_{\bar{\sigma}}^{r/a}(\omega_{\bar{\sigma}})}{\omega\!-\!%
\varepsilon_{d\sigma }\!-\! \Sigma^{\prime}_{1}(\omega_{\bar{\sigma}%
})\!\pm\! \mathrm{i}\,\Gamma (1\!+\!\tilde{f}(\omega_{\bar{\sigma}
}))\!\mp\! 2\mathrm{i}\,\Gamma P_{\bar{\sigma} }^{r/a}(\omega_{\bar{\sigma}})%
}\;,  \label{G}
\end{equation}
with $\omega_{\bar{\sigma}}=\omega+{\bar{\sigma}}g\mu_BB$ ($\bar{\sigma}\equiv -\sigma$), 
\begin{equation}
\Sigma^{\prime}_{1}(\omega )=\sum_{\alpha,k}|V_{\alpha k}|^{2}\frac{%
f_{\alpha }(\varepsilon _{k\sigma } )}{\omega-\varepsilon _{k\sigma }}\,,
\end{equation}
and 
\begin{equation}
P_{\sigma }^{r/a}(\omega ) =\frac{\Gamma}{\pi}\int_{-D}^D \tilde{f}(\omega ^{\prime }) \frac{G_{d\sigma }^{a/r}(\omega^{\prime})}{%
\omega\pm\mathrm{i}\,0^+ -\omega ^{\prime }}\,\mathrm{d}\omega'.  \label{P}
\end{equation}

Here, $\Gamma _{\alpha }=\pi \sum_{k}|V_{\alpha k}|^{2}\times \delta (\omega
-\varepsilon _{k\sigma })$ defines the coupling to the leads
for which we assume a constant density of states of $1/2D$ per spin, $\Gamma
=\sum_\alpha\Gamma_{\alpha}$ and $%
f_{\alpha}(\omega)=(\exp {[\beta (\omega\!-\!\mu_{\alpha}})]+1)^{-1}$  is the Fermi distribution function of the $\alpha$-lead with $%
\beta =1/k_{B}T$.  Notice that the set of equations (\ref{G}) and (\ref{P}), which have to be solved self-consistently using
numerical methods, have the same form as in the equilibrium case\cite{Kashcheyevs2006} but with an effective distribution function 
\begin{equation}
\tilde{f}%
(\omega )=\sum_\alpha\frac{\Gamma_{\alpha}}{\Gamma}f_\alpha(\omega). 
\end{equation}
The reason for this is that, within the decoupling scheme used for solving the EOM, the impurity's lesser Green function can only take the pseudo-equilibrium form $G_{d\sigma}^<(\omega )=\tilde{f}(\omega )\left[G_{d\sigma }^{a}(\omega )-G_{d\sigma }^{r}(\omega )\right]$---see the appendix.

\begin{figure}[t]
\includegraphics[width=.45\textwidth,clip]{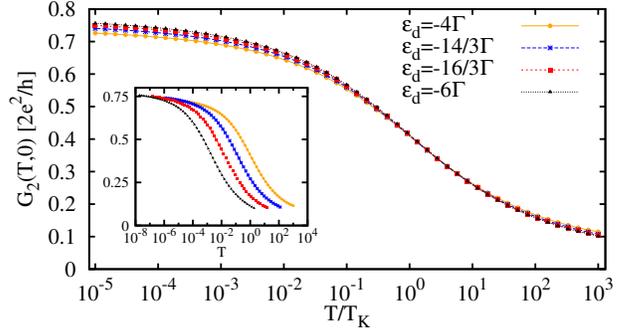}
\caption{(Color online) Linear conductance as a function of $T/T_K$ for
different values of $\varepsilon_d$ and for $\Gamma=0.015D$. There is
a very good scaling with the effective Kondo temperature $T_K=D\exp(\pi%
\varepsilon_d/\Gamma)$. Notice that the scaling is improved as $|%
\varepsilon_d|/\Gamma$ increases and that the exact quadratic low $T$
behavior of the conductance is not captured in this approach. Inset:
Unscaled conductance.}
\label{G-E}
\end{figure}

At equilibrium ($\tilde{f}(\omega )\equiv f(\omega )$), the EOM approach leads
to the impurity  density of states $\rho_\mathrm{eq}(\omega)$ shown in Fig. \ref{Ad-E}. It presents a clear
Kondo peak near  ($\omega=0$). A detail of the Kondo peak for different
temperatures is shown in the inset. As stated in Refs. [\onlinecite{Kashcheyevs2006,Entin-Wohlman2005}], as 
$T\!\rightarrow\!0$ the  impurity's spectral density satisfies
the Fermi liquid unitary condition, $\rho_\mathrm{eq}(\omega =0)=2/\pi\Gamma$. Here, however, that limit is reached only logarithmically 
as the exact low temperature $T^{2}$ dependence is not recovered within this approach.\cite{Entin-Wohlman2005}

In what follows we use these results to probe the scaling laws of the
transport properties in different configurations with and without an external
magnetic field $B$.

\section{Two terminal configuration and scaling laws\label{2t}}

The current $I$ flowing through the impurity in the two terminal setup can be written as 
\cite{Meir1992}
\begin{equation}
I=\frac{2e}{\hbar}\frac{\Gamma _{L}\Gamma _{R}}{\Gamma _{L}+\Gamma _{R}}\;\int %
\rho_d(\omega ) [ f_{L}(\omega )-f_{R}(\omega )]\, \mathrm{d}\omega\;.
\label{I}
\end{equation}
The two terminal linear conductance, given in terms of the equilibrium
density of states $\rho_\mathrm{eq}(\omega )$, is then 
\begin{equation}
G_2(T,0)=\frac{2e^{2}}{\hbar}\frac{\Gamma _{L}\Gamma _{R}}{\Gamma _{L}+\Gamma _{R}%
}\int \left(-\frac{\partial f}{\partial \omega}\right)\;\rho_%
\mathrm{eq}(\omega )\,\mathrm{d}\omega .  \label{G-equi}
\end{equation}
The results for $G_2(T,0)$ vs $T/T_{K}$ for four different values of $%
\varepsilon_d/\Gamma$ and $\Gamma_L=\Gamma_R$ are shown in Fig. \ref{G-E}.
Here, $T_{K}=D\exp{(\pi\varepsilon _{d}/\Gamma )}$ is the relevant energy
scale, or the effective Kondo temperature, that is obtained within this
approach.\cite{Kashcheyevs2006} Unscaled data is shown in the inset for comparison. For
all the parameters studied, we obtain an excellent scaling in the
temperature range $0.1T_{K}\lesssim T\lesssim 10T_{K}$. Note that for
large $|\varepsilon _{d}|/\Gamma $ the scaling extends over a wider range of
temperature. In addition, it is apparent from the figure the absence of the
characteristic $T^2$ Fermi liquid behavior expected at low $T$. Instead, a
logarithmic behavior is obtained.\cite{Entin-Wohlman2005}

\begin{figure}[t]
\includegraphics[width=.45\textwidth,clip]{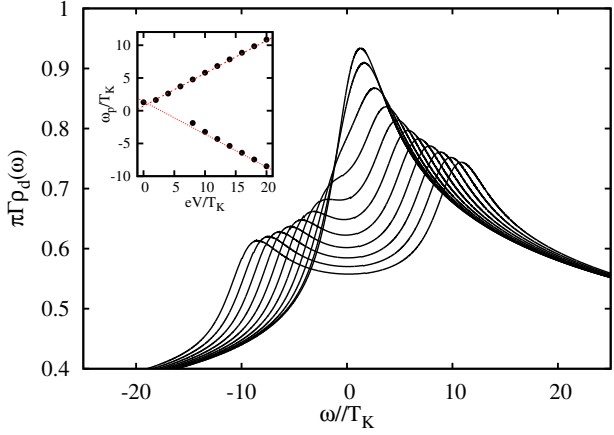}
\caption{(Color online) Non-equilibrium spectral density. The figure shows
the splitting of the Kondo peak for different values of the bias voltage, $%
eV\in[0,20T_K]$. Here $\varepsilon_d=-4\Gamma$ and $\Gamma=0.015D$.
Inset: Kondo peak position $\omega_\mathrm{p}$ as a function of
the bias. The dashed lines correspond to $ \omega_\mathrm{%
p}=\pm eV/2+ \omega_0^\pm$ where $ \omega_0^\pm$ is a fitting parameter.}
\label{Ad-noE}
\end{figure}

The non-equilibrium density of states $\rho_d(\omega)$ with $\mu _{L}=-\mu_R=eV/2$
is shown in Fig. \ref{Ad-noE} for several values of the bias voltage $V$ and 
$T=T_{K}$. As $V$ increases the Kondo peak splits with two maxima at
approximately $\omega_\mathrm{p}\simeq\pm eV/2$. This is shown in detail in the inset of the
figure, where the dashed lines, corresponding to the expression $\omega_%
\mathrm{p}=\pm eV/2+\omega_0^\pm$, are a guide to the eyes. As the Kondo peak has an asymmetric
lineshape, due to the background contribution to the spectral density, there
is a difference in the relative height of the two non-equilibrium peaks. It
is worth to point out that their heights will also change in the case of
asymmetry coupling, $\Gamma_L\neq\Gamma_R$.

\begin{figure}[t]
\includegraphics[width=.45\textwidth,clip]{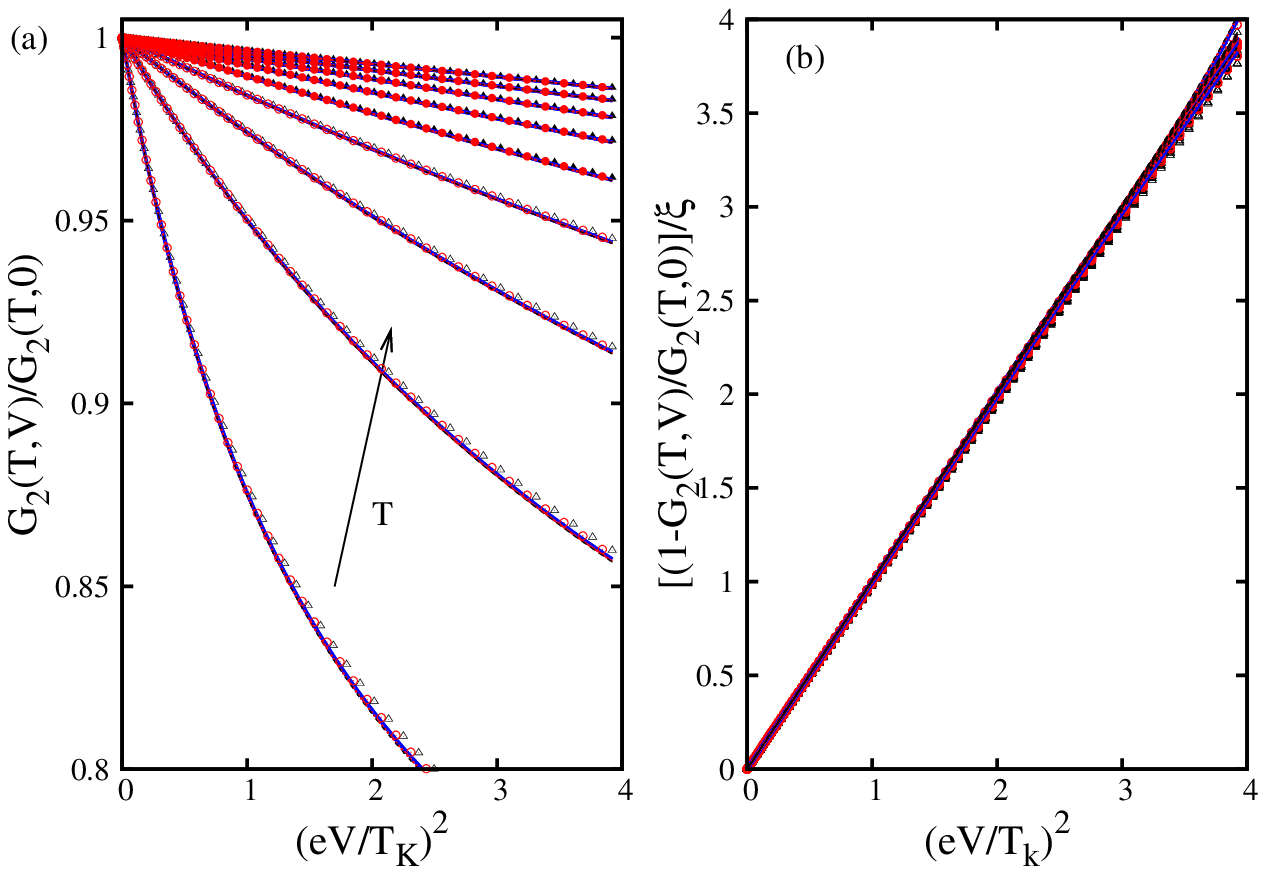} %
\includegraphics[width=.45\textwidth,clip]{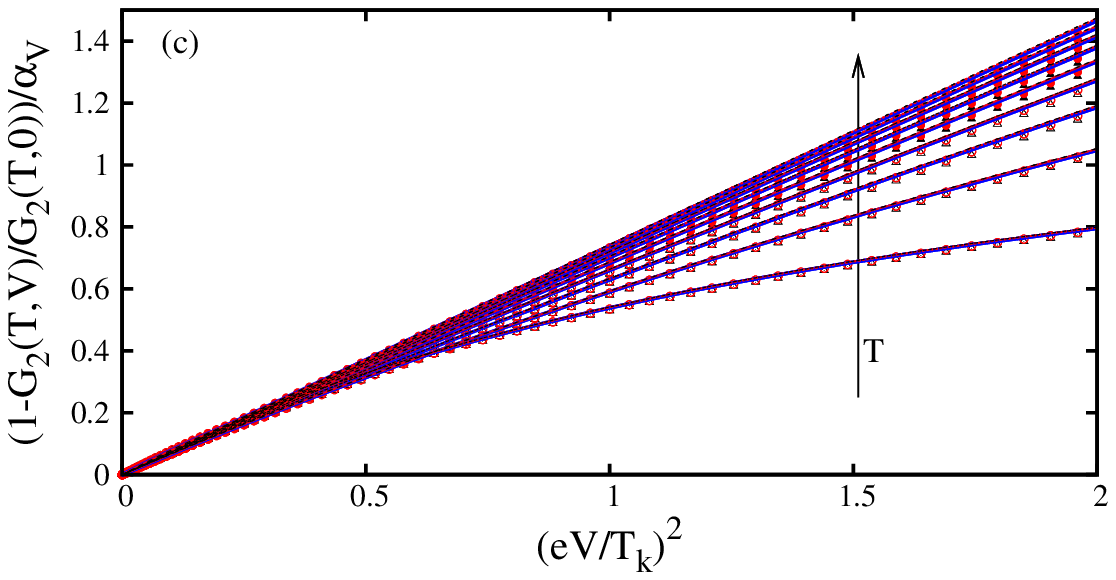}
\caption{(Color online) a) Normalized differential conductance as a function
of $(eV/T_K)^2$ for $T/T_K\in [0.25,2.25]$ and $ \varepsilon%
_d/\Gamma=-14/3$ triangles),$-16/3$ (circles),$-6$, $20/3$ and $7.07$ (lines). Open symbols
correspond to $T\leq T_K$ and the arrow indicates the direction of
increasing temperature. Note that the scaling with bias voltage is
reasonable good even beyond the quadratic regime.; b) Scaling of the
normalized differential conductance as discussed in the text. c) Scaling suggested in Ref. [\onlinecite{Grobis2008}] , Eq. (\ref{scaling-def}).}
\label{scaling}
\end{figure}

The differential conductance $G_2(T,V)=\partial I/\partial V$ is obtained as
the numerical derivative of Eq. (\ref{I}). The result, normalized by $%
G_2(T,0)$, is plotted in Fig. \ref{scaling}a) as a function of $%
(eV/T_{K})^{2}$ for different temperatures and five different values of $%
|\varepsilon_d|/\Gamma$. All the data corresponding to the same $T/T_K$ collapse into a single curve. This shows that
 the EOM approach is able to capture the universal scaling as a function of $eV/T_K$, as suggested in
Ref [\onlinecite{Lebanon2001}]. Furthermore, we are able to fit the numerical data by the following expression
\begin{equation}
 \frac{G_2(T,V)}{G_2(T,0)}=\sum_n a_n  \left(\frac{eV}{T_K}\right)^{2n}\,,
\label{expansion}
\end{equation}
where the coefficients $a_n$ are a function  of $T/T_K$ \textit{only}. This is shown in Fig. \ref{scaling}(b), 
where we have plotted $[1-G_2(T,V)/G_2(T,0)]/\xi$ as a function of $(eV/T_K)^{2}$ where $\xi=[a_1+a_2 (eV/T_K)^{2}+a_3(eV/T_K)^{4}+a_4 (eV/T_K)^{6}]$.
The good scaling shows that, within the EOM approach,  
\begin{equation}
 \frac{G_2(T,V)}{G_2(T,0)}=\mathcal{F}\left(\frac{T}{T_K},\frac{eV}{T_K}\right)
\end{equation}
 is indeed a universal function of $\textit{both}$ $eV/T_K$ \textit{and} $T/T_K$. 
 This is valid only in the $U\rightarrow\infty$ limit , as it is known that for finite $U$ the function $\mathcal{F}$ has non-universal contributions---it depends on the asymmetry of the couplings, for instance.\cite{Sela2009,Rincon2009} Finding the analytical form of $\mathcal{F}$ is, however, difficult.

The scaling properties of $G_2(T,V)$ were recently examinated by  experimental groups in both  quantum dots\cite{Grobis2008} and molecular 
systems.\cite{Scott2009} It was shown that the differential conductance approximately follows a rather
simple universal scaling relation, empirically chosen so that the low $T$
and low $V$ regime agrees with the quadratic behavior expected from Fermi
liquid theory. Namely, 
\begin{equation}
\left(1-\frac{G_2(T,V)}{G_2(T,0)}\right)= \alpha_{V} \left(\frac{eV}{T_{K}}%
\right)^{2}  \label{scaling-def}
\end{equation}
where $\alpha _{V}=c_{T}\alpha [1+c_{T}(\gamma /\alpha
-1)(T/T_{K})^{2}]^{-1}$---here $\alpha_V$ plays the role of $a_1$ in Eq. (\ref{expansion}). The parameters $c_{T}$, $\gamma $ and $\alpha $
depend on the nature of the system and experiments have found quite different values for quantum dots
or molecular systems.\cite{note0} While in GaAs quantum dots these parameters are in
reasonable agreement with theoretical calculations,\cite{Oguri2001,Rincon2009,Sela2009,Roura-Bas} in
molecular junctions there are systematic deviations from these values whose
origin is still unclear.\cite{note1}
Therefore, we use the experimental values obtained in Ref.[\onlinecite{Grobis2008}%
] to scale our numerical data for the purpose of comparison. The result is shown in Fig. \ref{scaling}(c). 
For $eV/T_{K}\lesssim 0.7$ all curves approximately
collapse into a single line. In agreement with the experimental observation
\cite{Grobis2008} and recent calculations,\cite{Rincon2009,Roura-Bas} deviations from the $%
(eV/T_{K})^{2}$ behavior are more pronounced for low values of $T/T_{K}$. This is related to the fact that the 
scaling function $\alpha_V$ is too simple to capture the universal scaling shown in Fig. \ref{scaling}(b). It 
is clear then, that a better test of universality would be to fit the experimental data with Eq. (\ref{expansion}) 
and check the scaling of the coefficients with $T/T_K$.

Several experimental groups have recently measured the effect of an in-plane magnetic
field on the differential conductance at finite bias voltages\cite{Kogan2004,Amasha2005,Quay2007,Liu2009} with 
different results. Since the Zeeman coupling shifts the impurity levels
generating a splitting $\Delta =g\mu _{B}B$ between the two spin states one would naively expect a splitting of 
the Kondo peak in the spectral density and a splitting of the zero bias anomaly in the differential conductance. From the theoretical perspective, this is indeed the situation for $\rho_\mathrm{eq}(\omega)$; it is well established that for sufficiently large $\Delta$
the Kondo peak splits in two spin resolved peaks at $\omega \simeq \pm
\Delta$. However, in order to measure  the field induced splitting in a two terminal setup, 
a finite bias voltage of the order of $\Delta$ needs to be applied. As we discussed above, this 
in turns leads to a bias induced splitting and therefore the experimental situation is more involved. 
In fact, in Ref. [\onlinecite{Quay2007}] the splitting of the zero bias anomaly was found to be 
equal to $2\Delta$ for large fields while in Refs. [\onlinecite{Kogan2004,Amasha2005,Liu2009}] it 
was found to be larger than $2\Delta$.

\begin{figure}[t]
\includegraphics[width=.45\textwidth,clip]{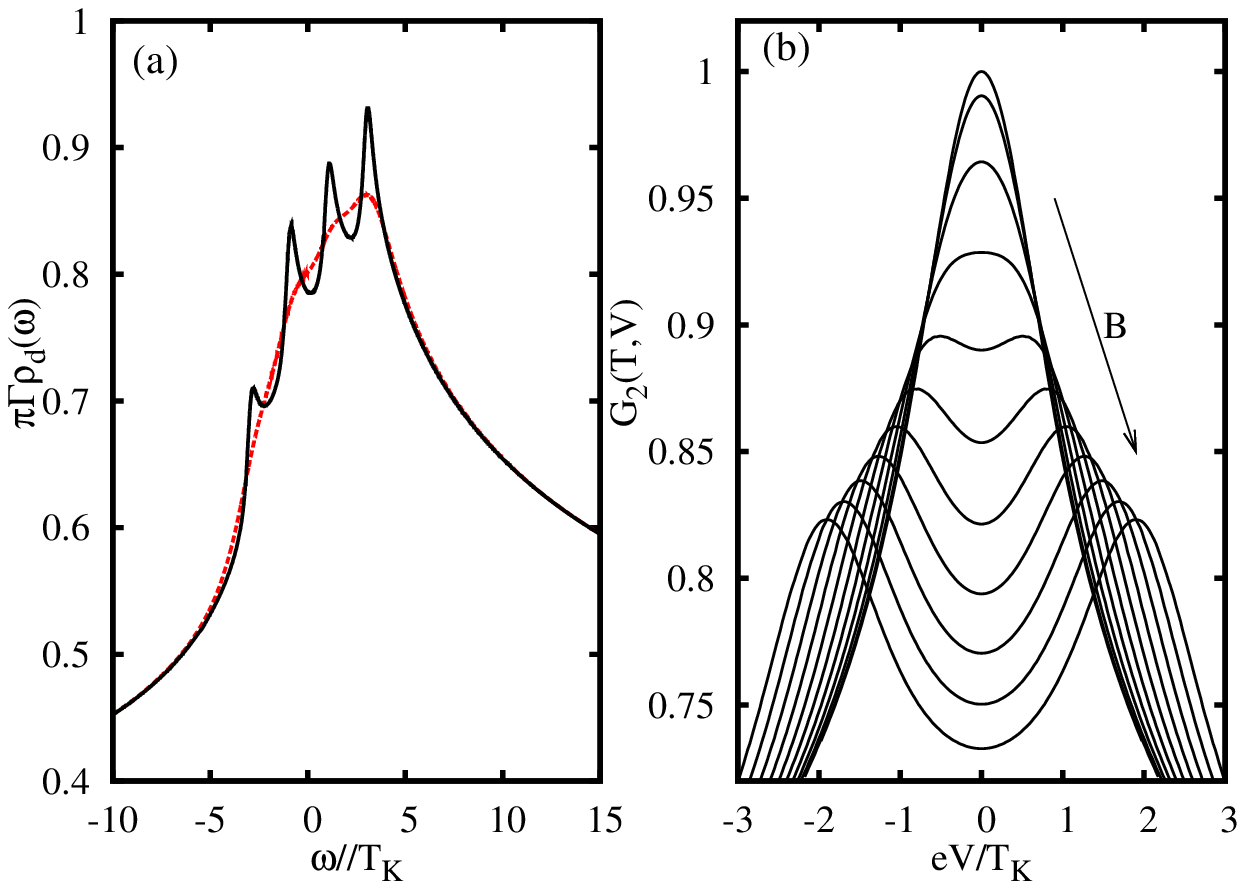}
\includegraphics[width=.45\textwidth,clip]{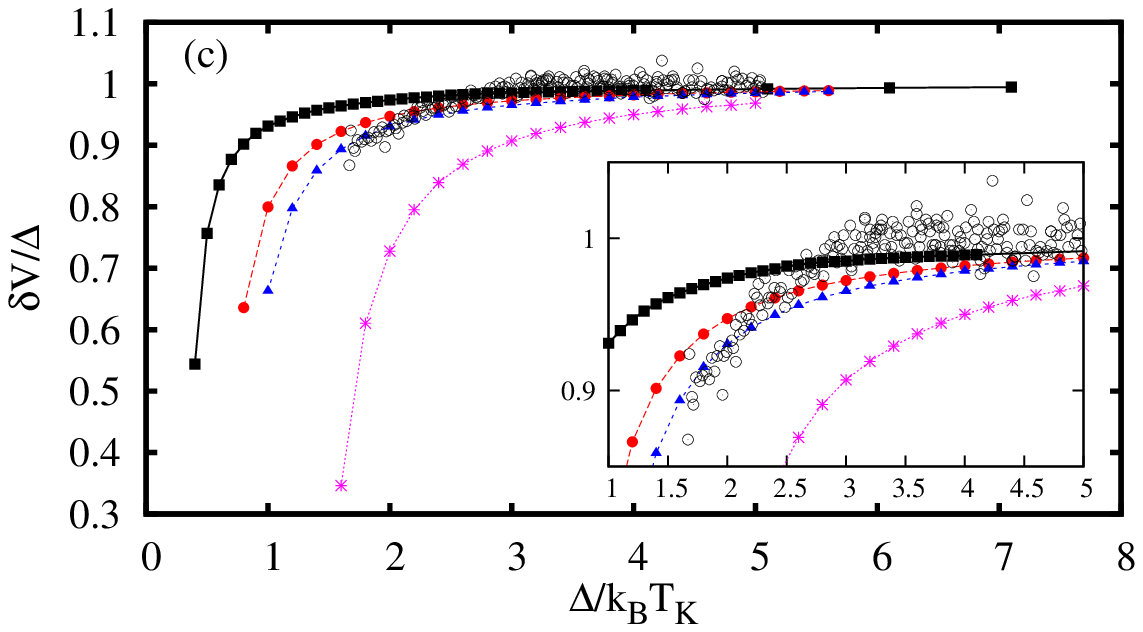}
\caption{(Color online) a) Non-equilibrium spectral
density for $\Delta/T_K=2$, $eV=\Delta$, $\varepsilon_d/\Gamma=16/3$ and $T/T_K=1/5$ (dashed line) and $T/T_K=1/10$
(solid line); b) Differential conductance for different values of
the Zeeman splitting, $\Delta/T_K\in [0,2]$ and $T/T_K=1/5$. The conductance is normalized by its zero bias and zero $B$ field value; c) Position of the conductance peak ($\pm\delta V$) as a function of the Zeeman
splitting for $T/T_K=1/2$ (\textasteriskcentered), $1/4$ ($\blacktriangle$), $1/5$ ($\bullet$) and $1/10$ ($\blacksquare$). The open dots are the experimental results of Ref. [\onlinecite{Quay2007}]. Note that the splitting of the conductance 
peak is well defined for $\Delta>0.4T_K$ when $T/T_K=1/10$. Inset: Zoom in around the experimental data.}
\label{GvsB}
\end{figure}

Figure \ref{GvsB}(a) shows the impurity spectral density calculated using the EOM approach for a finite bias 
and a finite Zeeman field ($eV=\Delta$). Two different temperatures and a symmetric drop of the bias voltage, $\mu_L=-\mu_R=eV/2$ are considered.
 For the lowest temperature, there are four peaks clearly visible at  $\omega\simeq\pm\Delta\pm eV/2$. The differential conductance $G_2(T,V)$, normalized to its $eV=\Delta=0$ value,  is shown in Fig. \ref{GvsB}(b) for $\Delta/T_K\in[0,2]$ and $T/T_K=1/5$. For this temperature, $G_2(T,V)$ presents a local
minimum at $eV=0$ and two maxima at $\pm \delta V$ for $\Delta/T_K\gtrsim0.8$. 
The splitting of the zero bias anomaly increases monotonically with $\Delta$ being clearly visible even though the corresponding peaks in the spectral density are barely observable.
The monotonic behavior of the conductance peak position $\delta V$ as a function of $\Delta/T_K$ is plotted in Fig. \ref{GvsB}(c) for different temperatures. The experimental data of Ref. [\onlinecite{Quay2007}], which corresponds to $T/T_K\sim1/6$, is included for comparison. The EOM results corresponding to $T/T_K\sim0.2$ are in qualitative agreement with those data. We note that $\delta V$ is well defined when $\Delta$ is greater than a certain critical value $\Delta_c$---for instance, $\Delta_c\approx0.4T_K$ for $T/T_K=1/10$---  which increases with $T$, as expected. We observe that for a large Zeeman field $\delta V\rightarrow\Delta$, in agreement with some previous theoretical estimates. We have verified (data not shown) that $\delta V$ depends only on $\Delta/T_K$, in contrast to the experimental data of Ref. [\onlinecite{Liu2009}].

It is worth to emphasize that even though $G_2(T,V)$ is not simply related to the impurity
spectral density, as in the linear regime, the value of the observed
splitting is already an indirect evidence of the strong bias dependence of $%
\rho_d(\omega)$. Indeed, had we calculated the differential conductance using the
equilibrium spectral density, Eq. (\ref{G-equi}), we would have found that $G_2(T\sim0,V)\propto \rho_\mathrm{eq}(eV/2)+\rho_\mathrm{eq}(-eV/2)$ and so
it would present, for large $\Delta$, two maxima at $\delta V\sim \pm 2\Delta $ rather than at $\pm
\Delta$.

\section{Three terminal configuration: probing the splitting of the Kondo
peak\label{3t}}

\begin{figure}[t]
\includegraphics[width=.45\textwidth,clip]{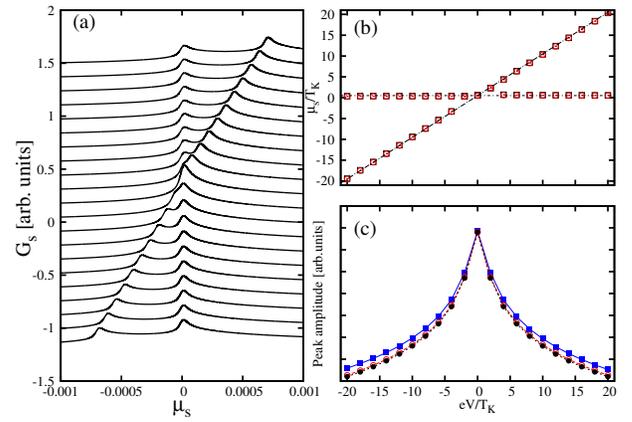}
\caption{(Color online) a) Scanning probe's differential conductance as a function of
 its chemical potential $ \mu_s$ for $ \mu_L=eV$, $%
 \mu_R=0$ with $eV$ taking different values in the range $%
[-20T_K,20T_K]$---the curves include an offset in the vertical axis for
clarity. The position of the conductance's peaks and the amplitude of the
highest are shown in b) and c), respectively, as a function of the bias
voltage for $ \varepsilon_d=-4\Gamma$($\bullet$), $-16/3\Gamma$($%
\blacksquare$) and $-6\Gamma$ ($\blacktriangle$). The peak amplitude shows a $\ln(eV/T_K)$ dependence.}
\label{fig7}
\end{figure}

As already mentioned, the differential conductance in the two lead configuration, $G_2(T,V)$, does not give a detailed
information of the voltage dependence of the impurity's spectral density $\rho_d(\omega )$. To scan the out of equilibrium Kondo resonance  an extra weakly coupled lead has to be attached to the impurity.\cite{Sun2001,Lebanon2001,Shah2006}
By sweeping the voltage of this extra lead it is possible to extract the
frequency dependence of the out of equilibrium  density of states $\rho_d(\omega )$. Here, we analyze the 
setups of Ref. [\onlinecite{DeFranceschi2002,Leturcq2005}] illustrated in
Fig.\ref{fig1}(b) and (c), respectively. The former corresponds to a dot coupled to an out of equilibrium wire. 
An electric current is flowing through the wire and a weakly coupled lead 
connected to the dot (with coupling constant $\Gamma_s$) acts as the scanning probe.

Assuming that the distribution functions of the left and right moving carriers in the  ballistic wire
are given by  $f_{L}(\omega )$ and $f_{R}(\omega )$, respectively, the current flowing from the scanning probe 
through the impurity can be effectively computed employing a three lead configuration as the one  presented 
 in Ref. [\onlinecite{Lebanon2001}].
When the scanning probe is biased to a voltage $V_{s}$, the current flowing
though the impurity is, to lowest order in $\Gamma _{s}$, \cite{Lebanon2001}
\begin{equation} \label{3l}
I_{s}=\frac{2e}{\hbar}\Gamma _{s}\int  \lbrack \tilde{f}%
(\omega )-f_{s}(\omega )]\;\rho_d(\omega )\, \mathrm{d}\omega\;,
\end{equation}%
where $f_{s}(\omega )=1/(\exp {\beta (\omega +eV_{s})+1)}$ is the Fermi
distribution of the probe, $\tilde{f}(\omega )$ is the (out of
equilibrium) carrier distribution in the  wire and $\rho_d$ is the impurity's  spectral density in the absence of the probe lead.
When the  impurity is coupled with the same coupling constant to left and right moving electrons, $\Gamma
_{L}=\Gamma _{R}$, and we have
 $\tilde{f}(\omega )= (f_{L}(\omega )+f_{R}(\omega ))/2  $. 
 The double step distribution  $\tilde{f%
}(\omega )$ is first used to evaluate the spectral density using Eqs.(\ref{G}) and (\ref{P})
 and then the current $I_{s}$. The differential conductance $%
G_{s}(T,V)=\partial I_{s}/\partial V_{s}$ is 
\begin{equation}
G_{s}(T,V)=-\frac{2e^{2}}{\hbar}\Gamma _{s}\int  \frac{%
\partial f_{s}(\omega )}{\partial V_{s}}\;\rho_d(\omega )\,\mathrm{d}\omega\,,  
\label{gs}
\end{equation}%
and at low temperatures $G_{s}(T,V)\approx (2e^{2}/\hbar)\Gamma _{s}\rho_d(eV_{s})$. In Fig. \ref{fig7} we present results 
for the differential conductance 
$G_{s}(T,V)$ as obtained from Eq. (\ref{gs}). We have set $\mu _{L}=eV$ and $%
\mu _{R}=0$. The scanning probe conductance as a function of the  voltage $V_{s}$
shows a double peak structure due to the splitting of the Kondo resonance.
As shown in Fig. \ref{fig7}(b), the conductance peaks are observed at $eV_{s}=\mu _{L}
$ and $eV_{s}=\mu _{R}$. The amplitude of the highest peak decreases
with the wire voltage $V$ and shows a good scaling when plotted as a
function of $eV/T_{K}$---in particular for $|\varepsilon_d|/\Gamma\gg1$. Once again, these results are in qualitative agreement with the experimental observations.\cite{DeFranceschi2002}
However, we obtain a clear $\ln(|eV|/T_K)$ dependence for the peak's amplitude in the range shown in the figure while the experimental data was fit to $A \ln^{-2}(|eV|/T_K)+ B  \ln^{-3}(|eV|/T_K)$.\cite{DeFranceschi2002} Nevertheless, due to the limited range and small values of $eV/T_K$ explore in the experiments, the $\ln(|eV|/T_K)$ dependence cannot be completely ruled out.
 
Similar results are obtained in the case of a single impurity attached to three leads when one of them is acting as a probe, see Fig.\ref{fig1}(c). In that case,  $\tilde{f}(\omega)= ( \Gamma _{L} f_{L}(\omega) + \Gamma _{R} f_{R}(\omega ))/ (\Gamma _{L}+\Gamma _{R}) $ and 
thus a left and right asymmetry in the couplings changes the
relative weights of the two splitted conductance peaks. In particular, for a large asymmetry  and in  the small bias regime, $eV<T_{K}$, the unsplitted peak tends to be pinned to the chemical potential of the strongest
coupled lead, as observed in Ref. [\onlinecite{Leturcq2005}]. 

\section{Summary\label{summary}}

We have discussed the out of equilibrium transport properties of an 
Anderson impurity embedded in a multi-terminal structure. In order to compute the relevant transport quantities
we have generalized  the EOM approach developed in Refs. [\onlinecite{Entin-Wohlman2005,Kashcheyevs2006}] to the non-equilibrium situation.
Within this approach, we have tested  the scaling laws recently reported in a series of experiments
 performed in quantum dots in contact to metallic leads or wires driven  out of equilibrium 
and/or in the presence of an external magnetic fields \cite{Kogan2004,Amasha2005,Grobis2008,Liu2009}.
We found that, within the EOM approach in the $U\rightarrow\infty$ limit,  ${G_2(T,V)}/{G_2(T,0)}$ is a universal function of 
$\textit{both}$ $eV/T_K$ \textit{and} $T/T_K$. 
Indeed, we have been able to fit the numerical data  by the expression shown in Eq. (\ref{expansion}), where the expansion coefficients
 $a_n$ are a function only of $T/T_K$. We then conclude that a fit of the experimental data to Eq. (\ref{expansion})  
instead of the simple  scaling function  $\alpha_V$,\cite{Grobis2008} would be a better test of universality.

In the presence of a Zeeman field we have found a monotonic behavior of the differential conductance peak position $\delta V$ as a function 
of $\Delta/T_K$, in agreement with the experimental data reported in Ref. [\onlinecite{Quay2007}]. 
In addition, for large Zeeman field we have found that $\delta V\rightarrow\Delta$  and that 
 $\delta V$ depends only on $\Delta/T_K$, in contrast to the experimental data of Ref. [\onlinecite{Liu2009}].

In order to scan the out of equilibrium impurity's spectral density  recently analyzed in a series of experiments
\cite{DeFranceschi2002,Leturcq2005}  we have  considered, besides the traditional two leads set up,
 additional configurations, namely an impurity embedded 
in a mesoscopic wire and the three leads configuration in which the extra lead acts as a weakly coupled 
scanning probe.  
The computed  scanning probe  conductance as a function of the tip voltage $V_{s}$
shows a double peak structure  due to the splitting of the Kondo resonance.
 The amplitude of the highest peak decreases as a function of the wire's bias with a $\ln(eV/T_{K})$ dependence. 
In summary, the formalism developed in this work  provides   a   reasonable description of
the out of equilibrium  physics of   a quantum impurity embedded in a rather general mesoscopic setup.
More involved setups and possible further  generalizations are currently under consideration.

\begin{acknowledgments}
We thanks A. Aligia for useful discussions.
We acknowledge financial support from PICT 06-483 from ANPCyT and PIPs 5254 and 11220080101821 
from CONICET, Argentina.

{\it Note added}: After completion of this work we learn of two recent manuscripts, Refs. [\onlinecite{lim,Roermund2010}], 
that employ a similar decoupling scheme of the EOM  to compute the impurity retarded Green function in the out of equilibrium
situation. 

\end{acknowledgments}

\appendix*

\section{Calculation of $G^r_d( \omega)$ using EOM}

In this appendix we present an extension of the EOM results of Ref.[\onlinecite{lacroix1981,Kashcheyevs2006,Entin-Wohlman2005}] for the out of equilibrium case.
We use the following standard definitions 
\begin{eqnarray}
\nonumber
\langle \langle A(t),B(t^{\prime })\rangle \rangle ^{r} &=&-\ci\theta
(t-t^{\prime })\langle A(t)B(t^{\prime })+B(t^{\prime })A(t)\rangle \\
\nonumber
\langle \langle A(t),B(t^{\prime })\rangle \rangle ^{a} &=&\ci\theta
(t^{\prime }-t)\langle A(t)B(t^{\prime })+B(t^{\prime })A(t)\rangle\,, \\
\end{eqnarray}
where the $a$ and $r$ supraindex stand for the advanced and retarded Green
functions, respectively. For the local Green functions in the stationary
state we use the notation $G_{d}^{r/a}(t-t^{\prime })\equiv \langle \langle
d_{\sigma }(t),d_{\sigma }^{\dag }(t^{\prime })\rangle \rangle ^{r/a}$ while $G_{d\sigma}^{r/a}(\omega )=\langle \langle
d_{\sigma },d_{\sigma }^{\dag }\rangle \rangle ^{r/a}$ indicates its Fourier transform (the same applies to a generic green function $\langle \langle A,B\rangle \rangle ^{r/a}$). We also define the
local lesser functions as: $G_{d\sigma}^{<}(t-t^{\prime })=-\ci\langle d_{\sigma
}^{\dag }(t^{\prime })d_{\sigma }(t)\rangle $ and its Fourier transform $%
G_{d\sigma}^{<}(\omega )$.

We now consider a local  impurity coupled to an arbitrary number of reservoirs.
Following the standard EOM procedure in the infinite $U$ limit, we obtain the following expression for the retarded and advanced local Green functions 
\begin{widetext}
\begin{eqnarray}
(\omega -\varepsilon _{d\sigma})G_{d\sigma}(\omega ) &=&1-\langle n_{\overline{\sigma }}\rangle -
\sum_{\alpha k}V_{\alpha }g_{\alpha k \overline{\sigma}}(\omega_{\bar{\sigma}} )\langle 
d_{\overline{\sigma }}^{\dag }c_{\alpha k,\overline{\sigma }}\rangle 
+\sum_{\alpha k,\beta q}V_{\alpha }V_{\beta }g_{\alpha k \overline{\sigma}}(\omega_{\bar{\sigma}} )[\langle
\langle d_{\overline{\sigma }}^{\dag }c_{\beta q,\overline{\sigma }%
}c_{\alpha k,\sigma },d_{\sigma }^{\dag }\rangle \rangle -\langle \langle
c_{\beta q,\overline{\sigma }}^{\dag }d_{\overline{\sigma }}c_{\alpha
k,\sigma },d_{\sigma }^{\dag }\rangle \rangle  \nonumber \\
&&+\langle \langle d_{\overline{\sigma }}^{\dag }c_{\alpha k,\overline{%
\sigma }}c_{\beta q,\sigma },d_{\sigma }^{\dag }\rangle \rangle -\langle
\langle c_{\beta q,\overline{\sigma }}^{\dag }c_{\alpha k,\overline{\sigma }%
}d_{\sigma },d_{\sigma }^{\dag }\rangle \rangle ]  
\end{eqnarray}
\end{widetext}
where $\langle ..\rangle $ indicates the (out of equilibrium) mean value, $n_{\sigma }=d_{\sigma
}^{\dag }d_{\sigma }$, $\alpha $ and $\beta $ label the reservoir and 
$g_{\alpha k\sigma}(\omega )=(\omega -\varepsilon _{\alpha k \sigma})^{-1}$ is the
reservoir propagator and $\omega_{\bar{\sigma}} = \omega + \bar{\sigma} g\mu_B B$, where  $B$ is the external Zeeman field.
The decoupling scheme proposed by Lacroix \cite{lacroix1981} leads to the following
expression 
\begin{equation}
\left[{\mathcal{G}_{\sigma}}^{r/a}(\omega )\right]^{-1}G_{d\sigma}^{r/a}(\omega)=
1-\langle n_{\bar{\sigma}} \rangle -P_{\bar{\sigma}}^{r/a}(\omega_{\bar{\sigma}}) \;
\end{equation}
with 
\begin{equation}
\left[\mathcal{G}_{\sigma}^{r/a}(\omega )\right]^{-1}=(\omega\pm\ci 0^+ -\varepsilon _{d\sigma}-
\Sigma^{r/a} _{0 \sigma}(\omega )-\Sigma _{\bar{\sigma}}^{r/a} (\omega_{\bar{\sigma}}) )\;.
\end{equation}
$\Sigma^{r/a} _{ \sigma}(\omega )$ can be written as the sum of three different contributions,
 $\Sigma^{r/a} _{\sigma}(\omega )=\Sigma^{r/a} _{1\sigma}(\omega )+
\Sigma^{r/a} _{2\sigma}(\omega) - P_{\sigma}^{r/a}(\omega ) 
\Sigma^{r/a} _{0\bar{\sigma}}(\omega_{\bar{\sigma}} )$, where
\begin{equation}
\Sigma^{r/a}_{0\sigma}(\omega) \smeq\sum_{\alpha k}V_{\alpha }^{2}g^{r/a}_{\alpha k\sigma}(\omega )
\end{equation}
is the non-interacting self-energy and 
\begin{equation}
\Sigma^{r/a} _{1\sigma}(\omega ) \smeq\sum_{\alpha k}V_{\alpha }^{2}
g^{r/a}_{\alpha k\sigma}(\omega)f_{\alpha }(\varepsilon _{\alpha k \sigma} )  
\end{equation}

\begin{equation}
\nonumber
\Sigma^{r/a} _{2\sigma}(\omega)\smeq\!\sum_{\alpha k,\beta q}V_{\alpha }V_{\beta }\langle
c_{\beta q,\sigma}^{\dag }c_{\alpha k,\sigma}\rangle
g^{r/a}_{\alpha k\sigma}(\omega )\!-\!\Sigma^{r/a} _{1\sigma}(\omega )\,.  \\
\end{equation}
Here $f_{\alpha }(\omega )$ is the Fermi function of lead $\alpha $, and
\begin{eqnarray}
P_{\sigma}^{r/a}(\omega )&=&\sum_{\alpha k}V_{\alpha }g^{r/a}_{\alpha k\sigma}(\omega) 
 \langle d_{\sigma}^{\dag }c_{\alpha k,\sigma}\rangle \; .\label{pr}  
\end{eqnarray}

As mentioned above, all mean values
need to be evaluated in the out of equilibrium situation. In our case, the two relevant mean values are
\begin{equation}
\langle d_{\sigma}^{\dag }c_{\alpha k,\sigma}\rangle
\smeq-\ci V_{\alpha }\!\!\int\! \frac{\mathrm{d}\omega}{2\pi }\!\left[g_{\alpha k\sigma}^{<}(\omega)G_{d\sigma}^{a}(\omega)\smpl g_{\alpha k\sigma}^{r}(\omega )G_{d\sigma}^{<}(\omega)\right]\label{mean1}
\end{equation}%
and
\begin{widetext}
\begin{eqnarray}
\nonumber
\langle c_{\beta q,\sigma}^{\dag }c_{\alpha k,\sigma}\rangle  &=&\delta _{\alpha k,\beta q}f_{\alpha }(\varepsilon _{\alpha
k\sigma})
\\&&
-\ci V_{\alpha }V_{\beta }
\!\int\! \frac{\mathrm{d}\omega }{2\pi }\left[g_{\alpha k\sigma}^{<}(\omega )G_{d\sigma}^{a}(\omega )g_{\beta q\sigma}^{a}(\omega )+g_{\alpha
k\sigma}^{r}(\omega )G_{d\sigma}^{<}(\omega )g_{\beta q\sigma}^{a}(\omega
)+g_{\alpha k\sigma}^{r}(\omega )G_{d\sigma}^{r}(\omega )g_{\beta q\sigma}^{<}(\omega )\right]\,,\nonumber\\\label{mean2}
\end{eqnarray}
\end{widetext}
where $g_{\alpha k\sigma}^{<}(\omega )=2\pi \ci f_{\alpha }(\omega)\delta (\omega-\varepsilon _{\alpha k\sigma})$ is the non-interacting lesser Green function
of the $\alpha-$lead. Then, the calculation of these mean values requires the knowledge of the impurity's lesser green function $G_{d\sigma}^{<}(\omega)$. An equation for  $G_{d\sigma}^{<}(\omega)$ could be obtained by analytic continuation of the EOM equation if the \textit{matrix order} is preserved. However, in this case it is possible to obtain $G_{d\sigma}^{<}(\omega)$ by simply imposing that the impurity's Green function must satisfy $G_{d\sigma}^{r}=G_{d\sigma}^{a*}$ where $^*$ indicates complex conjugation. Indeed, this condition implies that
\begin{equation}
 P^r_{\bar{\sigma}}(\omega_{\bar{\sigma}})-P^{a*}_{\bar{\sigma}}(\omega_{\bar{\sigma}})=G_{d\sigma}^r(\omega)\left[\Sigma_{\bar{\sigma}}^r(\omega_{\bar{\sigma}})-\Sigma_{\bar{\sigma}}^{a^*}(\omega_{\bar{\sigma}})\right]\,.
\label{cond}
\end{equation}
From the definition of $P^r_{\sigma}(\omega)$, it is possible to show that 
\begin{equation}
 P^r_\sigma(\omega)-P^{a*}_\sigma(\omega)\smeq- \int \frac{\mathrm{d}\omega}{\pi}\frac{\mathcal{H}_\sigma(\omega')}{\omega-\omega'}+\ci \mathcal{H}_\sigma(\omega)\label{prpa}
\end{equation}
where 
\begin{equation}
\mathcal{H}_\sigma(\omega)=\Gamma_\sigma(\omega)\left(G_{d\sigma}^<(\omega)+\tilde{f}_\sigma(\omega)\left[G_{d\sigma}^r(\omega)-G_{d\sigma}^a(\omega)\right]\right)\,,
\end{equation}
and
\begin{equation}
 \tilde{f}_\sigma(\omega)=\sum_\alpha \frac{\Gamma_{\alpha\sigma}(\omega)}{\Gamma_\sigma(\omega)}f_\alpha(\omega)\,.
\end{equation}
Here, $\Gamma_\sigma(\omega)=\pi\sum_\alpha V_\alpha^2\rho_{\alpha\sigma}(\omega)=\sum_\alpha \Gamma_{\alpha\sigma}(\omega)$ with $\rho_{\alpha\sigma}$ the spin dependent density of states of the $\alpha$-lead. A similar, but cumbersome, expression can be obtained for 
$\Sigma_{\sigma}^r(\omega)-\Sigma_{\sigma}^{a^*}(\omega)$. The important point to emphasize is that, as in the case of Eq. (\ref{prpa}), $\Sigma_{\sigma}^r(\omega)-\Sigma_{\sigma}^{a^*}(\omega)$ is proportional to $\mathcal{H}_\sigma$. Therefore, $\mathcal{H}_\sigma(\omega)\equiv0$ is \textit{always} a consistent solution of Eq. (\ref{cond}). Probing that it is the only solution in the general case is not straightforward. However, in the absence of an external magnetic field or in the case that the Zeeman splitting is the same for the impurity and the band states, $\Gamma_{\bar{\sigma}}(\omega_{\bar{\sigma}})=\Gamma_{\sigma}(\omega)$, one obtains that
\begin{equation}
 \Sigma_\sigma^r(\omega)-\Sigma_\sigma^{a^*}(\omega)\smeq- \Gamma_{\bar{\sigma}}(\omega_\sigma) \mathcal{H}_\sigma(\omega)
\end{equation}
Therefore, if we assume that $\mathcal{H}(\omega)\ne 0$, we arrive to the conclusion that $-\mathrm{Im}G_{d\sigma}^r(\omega)=1/\Gamma_\sigma(\omega)$, which is clearly an unphysical result---note that $\mathcal{H}_\sigma^*(\omega)=-\mathcal{H}_\sigma(\omega)$. Then, we conclude that  $\mathcal{H}_\sigma(\omega)\equiv0$, which corresponds to
\begin{equation}
 G_{d\sigma}^<(\omega)=-\tilde{f}_\sigma(\omega)\left[G_{d\sigma}^r(\omega)-G_{d\sigma}^a(\omega)\right]\label{g<}\,.
\end{equation}
This result is nothing but the Ng's ansatz,\cite{Ng1996} which at the this level of decoupling of the EOM is the only consistent solution for $G_{d\sigma}^<(\omega)$. Inserting Eq. (\ref{g<}) into Eqs. (\ref{mean1}) y (\ref{mean2}), we obtain a set of self-consistent equations  for $G_{d\sigma}^r(\omega)$. In the  wide band limit, $\Gamma_{\alpha\sigma}(\omega)=\Gamma_{\alpha}$, we get
\begin{equation}
P^{r/a}_\sigma(\omega ) =\frac{\Gamma}{\pi}\int  \frac{\tilde{f}(\omega ^{\prime })G_{d\sigma }^{a/r}(\omega^{\prime})}{%
\omega\pm\mathrm{i}\,0^+ -\omega ^{\prime }}\,\mathrm{d}\omega'.  
\end{equation}
Similarly, we have $\Sigma _{2\sigma}^{r/a}(\omega )=\pm \ci\Gamma P^{r/a}_\sigma(\omega )$.
Collecting all these expressions we finally obtain 
\begin{equation}
G_{d\sigma}^{r}(\omega )\smeq\frac{1-\langle n_{\bar{\sigma }}\rangle
-P^{r}_{\bar{\sigma}}(\omega_{\bar{\sigma}} )}{\omega \smmi\varepsilon _{d\sigma}\smmi\mathrm{Re}\Sigma _{1\bar{\sigma}}(\omega_{\bar{\sigma}} )\smpl
\ci\Gamma (1\smpl\tilde{f}(\omega_{\bar{\sigma}} ))\smmi 2\ci\Gamma P^{r}_{\bar{\sigma}}(\omega_{\bar{\sigma}} )}\;.
\end{equation}

\end{document}